# Structural Inhomogeneities and Suppressed Magneto-Structural Coupling in Mn-Substituted GeCo$_2$O$_4$

Shivani Sharma[1*], Pooja Jain[2], Benny Schundelmier[1], Chin-Wei Wang[3], Poonam Yadav[4], Adrienn Maria Szucs[1], Kaya Wei[1], N. P. Lalla[2], Theo Siegrist[1,4]

[1] *National High Magnetic Field Laboratory, Tallahassee, FL 32310, United States.*
[2] *UGC-DAE Consortium for Scientific Research, Indore, Khandwa Road, MP, India.*
[3] *National Synchrotron Radiation Research Center, Hsinchu 300092, Taiwan.*
[4] *FAMU-FSU College of Engineering, Tallahassee, FL 32310, United States*
[4] *Center for Integrated Nanostructure Physics, Institute for Basic Science, Suwon 16419, Republic of Korea.*
*Corresponding author's email: phy.shivanisharma@gmail.com

**Abstract**

A comprehensive study of Ge$_{1-x}$Mn$_x$Co$_2$O$_4$ (GMCO) system was conducted using neutron powder diffraction (NPD), x-ray diffraction (XRD), Scanning electron microscopy, magnetometry, and heat capacity measurements. Comparative analysis with GeCo$_2$O$_4$ (GCO) highlights the influence of Mn substitution on the crystal and magnetic structure at low temperature. Surprisingly, phase separation is observed in GMCO with a targeted nominal composition of Ge$_{0.5}$Mn$_{0.5}$Co$_2$O$_4$. SEM/EDX analysis reveals that the sample predominantly consists of a Mn-rich primary phase with approximate stoichiometry Mn$_{0.74}$Ge$_{0.18}$Co$_2$O$_4$, along with a minor Ge-rich secondary phase of composition Ge$_{0.91}$Mn$_{0.19}$Co$_2$O$_4$. Although both GCO and GMCO crystallize in cubic symmetry at room temperature, a substantial difference in low-temperature structural properties has been observed. Magnetic and heat capacity data indicate ferrimagnetic ordering in the Mn-rich phase near $T_C$ = 108 K, while the Ge-rich phase exhibits antiferromagnetic order at $T_N$ = 22 K in GMCO. Analysis of heat capacity data reveals that the estimated magnetic entropy amounts to only 63% of the theoretical value expected in GMCO. A collinear ferrimagnetic arrangement is observed in the Mn rich phase below the magnetic ordering temperature, characterized by antiparallel spins of the Mn at A site and Co at B site along the c-direction. At 5 K, the refined magnetic moments are 2.31(3) for Mn$_A$ and 1.82(3) $\mu_B$ for Co$_B$ in the Mn rich ferrimagnetic phase. The magnetic structure at 5 K in the Ge rich secondary phase is identical to the antiferromagnetic structure of the parent compound GeCo$_2$O$_4$. The refined value of the Co$_B$ moment in this phase at 5 K is 2.53(3) $\mu_B$.

**Introduction**

Over the past several decades, the exploration of frustration in magnetic systems has been a subject of intense interest, primarily driven by the captivating magnetic states, such as spin glass or spin liquid in strongly correlated electron systems [1–10]. One of the most studied origins of frustration in magnetic systems is the geometrical arrangement of the first nearest neighbour antiferromagnetic (AFM) interactions in triangular (2D), tetrahedral (3D), pyrochlore, or Kagome lattices[5,10–14]. This discussion particularly focuses on geometrically frustrated systems, exemplified by the Co-based cubic spinel structures (ACo$_2$O$_4$), where the Co atoms are arranged in alternate planes of triangular and Kagome lattice, stacked alternatively along [111] direction as shown in figure 1[3,4,13,15–20]. Due to the corner-sharing tetrahedra of B-site cations (see the figure 1a) forming the pyrochlore lattice, the cubic spinel exhibits geometrical frustration. Frustration in spinel compounds can be attributed to the Jahn-Teller (JT) effect and spin-orbit interactions, leading to phenomena like orbital glass and liquid states[8,18,21–23]. Moreover, spinel oxides, distinguished by different cation distributions, are broadly classified into two categories: Normal Spinel and Inverse Spinel. In a normal spinel, A cations exclusively occupy tetrahedral sites, and B cations occupy octahedral sites. Each formula unit features eight tetrahedral and four octahedral sites. A notable example of a normal spinel is $GeCo_2O_4$ (GCO)[19]. However, in the inverse spinel configuration, all A cations and half of the B cations occupy octahedral sites, while the remaining B cations occupy tetrahedral sites, such as the case in $MnCo_2O_4$ (MCO)[24,25].

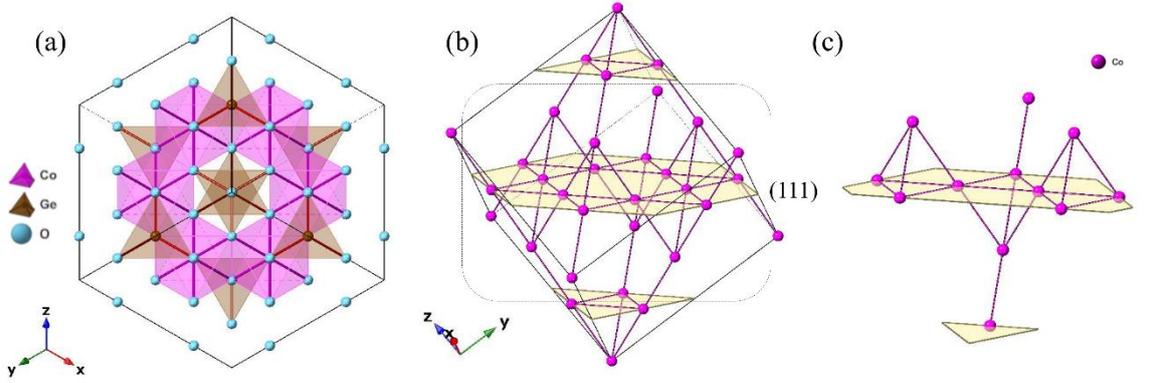

Figure 1: (a) Crystal structure of the spinel compound projected along the [111] direction, highlighting the network of corner-sharing $CoO_6$ (pink octahedra) and $GeO_4$ (brown tetrahedra) units within the cubic oxygen (blue spheres) framework. (b) Three-dimensional view of the Co sublattice emphasizing the pyrochlore network, with the (111) plane highlighted. (c) Zoomed-in view of the Kagome-triangular stacking motif in the Co sublattice along the [111] direction, illustrating the alternating arrangement of Kagome and triangular layers characteristic of geometrically frustrated systems.

Significant research has been conducted on both GCO and MCO, shedding light on their structural characteristics and magnetic properties. GCO has garnered significant attention due to its unique electronic and magnetic ground state featuring octahedral $Co^{2+}$. This state is characterized by a high spin $3d^7$ configuration with S=3/2, L=3, yet is more accurately described as a Kramer's doublet with $J_{eff}$=1/2. The presence of orbitally degenerate $t_{2g}^5$ states in the high-spin octahedral $Co^{2+}$ leads to substantial spin-orbit coupling, resulting in a pronounced single-ion anisotropy, a distinctive trait for a 3$d$ transition metal. Below its Néel temperature ($T_N \cong 21$ K), GCO exhibits AFM ordering with a characteristic wave vector k = (1/2,1/2,1/2) [3,4,11–13,15,19,21,26]. This ordering is accompanied by a structural phase transition from cubic to tetragonal symmetry[19]. Notably, neutron studies conducted by multiple research groups offer a cohesive understanding of the spin ordering in GCO[3,12,21,26]. On the other hand, MCO has been studied for its remarkable magnetic properties and colossal magnetoresistance behaviour [24,25]. The compound MCO is notably intriguing, particularly below 130 K, where magnetic hysteresis curves exhibit unconventional behaviour, which was suggested to be linked with the irreversible domain wall movements that overcome the anisotropy field below 130 K[24]. However, the unusual magnetic hysteresis as observed in the case of MCO warrants further investigations to exclude the artifacts including the trap field effect on the magnetic response on a single crystalline sample. Due to the interesting physical and structural properties of these compounds and the possible tuning of these properties with substitutions, several detailed studies have been performed on A and/or B site substituted spinel. The first NPD studies on $Co_2Ru_{1-x}Mn_xO_4$ provide evidence of change in magnetic order with Ru substitution and report two transitions at 100 and 180 K[25]. For lower Ru contents, a coexistence of long-range and short-range magnetic order was found where the complete Mn dilution results in the spin-glass-like ordering with a spin-freezing temperature of 16 K[25,27]. Further, in Bi-substituted MCO, the magnetic transition increases significantly to 200 K[28]. Moreover, in a recent study by Pramanik *et al.*, the ferrimagnetic ordering was confirmed at 184 K and 164 K in off-stoichiometric $Mn_{1.15}Co_{1.85}O_4$ and $Mn_{1.17}Co_{1.60}Cu_{0.23}O_4$[29]. In addition, Pramanik *et al.* have confirmed the ferrimagnetic ground state of $Ti_{0.6}Mn_{0.4}Co_2O_4$ and $Ti_{0.8}Mn_{0.2}Co_2O_4$ below 110.3 K and 78.2 K, respectively [30]. While prior studies have extensively analyzed the properties of the end members, the specific structural and magnetic transformations triggered by varying degrees of Mn substitution at the Ge site remain largely unexplored. A recent report with 20% Mn substitution at the Ge site based primarily on bulk magnetization measurements provides a limited understanding of the magnetic and structural properties at low temperatures [32].

A comprehensive study of $Ge_{1-x}Mn_xCo_2O_4$ (GMCO) was performed using a combination of neutron powder diffraction (NPD), x-ray diffraction (XRD), Scanning electron microscopy/EDX, magnetometry, and heat capacity techniques. The end members GCO and MCO have cubic and tetragonal symmetry at room temperature, respectively[6,19,24] due to the Jahn-teller active $Mn^{3+}$ ion in MCO. Therefore, the Mn substitution is anticipated to change the crystal symmetry and related geometrical frustration, which has a significant impact on magnetic exchange interactions. As a result, it is expected that the Mn substitution will not only alter the structural properties but will also affect the magnetic ground state, which is the main motivation for this work. Comparative analysis of GMCO with the parent compound $GeCo_2O_4$ (GCO) reveals the effects of Mn substitution on the crystal and magnetic structure, particularly at low temperatures. SEM/EDX analysis indicates clear phase separation, with a Mn-rich primary phase and a minor Ge-rich secondary phase. Magnetic measurements show ferrimagnetic ordering in the Mn-rich phase at ~108 K and AFM ordering in the Ge-rich phase at ~22 K, with refined magnetic moments consistent with expected values.

**Experimental**

Polycrystalline samples of $Ge_xMn_{1-x}Co_2O_4$ (x= 0 and 0.5) were prepared by a solid-state reaction method[4,15,18,21]. The starting materials, $GeO_2$ (99.99% purity), $MnO_2$ (99.99% purity), and $Co_2O_3$ (99.99% purity) were mixed in the stoichiometric amounts. The resulting mixture was then calcined at 950 °C for 12 hr and sintered at 950 °C for another 12 hr. To verify the phase purity of the prepared compounds and to investigate the low-temperature structural phase transitions, powder X-ray diffraction (XRD) measurement was recorded as a function of temperature using Rigaku's diffractometer equipped with Cu-Kα radiation. High-resolution imaging, Backscattered Electron X-ray (BEX), and Energy-Dispersive X-ray (EDX) analysis were performed using a Tescan MIRA 4 field emission scanning electron microscope (FE-SEM) equipped with an Oxford Instruments Unity BEX Imaging Detector and Ultim Max 170 mm² EDX detectors. BEX imaging was utilized to examine the raw powder sample, providing both topographical and compositional contrast, enabling the observation of chemical variations. For EDX analysis, the sample was epoxy-mounted and polished. For EDX analysis, a beam current of 1 nA was used at 20 kV. BEX imaging was conducted at 20 kV with a beam current of 3 nA, with an image scan and X-ray scan size of 1024, a dwell time of 400 microseconds, and an overall frame time of 419.4 seconds. Data processing and analysis were performed using Oxford Instruments Nano-Analysis AZtec software. The DC magnetic susceptibility measurements were measured using a superconducting quantum interference device (SQUID) as a function of temperature and field, and the heat capacity was measured using a physical property measurement system. Powder neutron diffraction measurements on GMCO were carried out at ECHIDNA and WOMABT beamlines at the OPAL facility, ANSTO,

Australia. The room temperature data from the ECHIDNA beamline with neutron wavelength 1.62 Å was used to estimate the chemical composition of the Mn substituted compound. Temperature-dependent data from WOMBAT beamline with neutron wavelength 2.41 Å was used for magnetic structure analysis as a function of temperature for a broad range of temperatures ranging from 5 to 120 K. Fullprof suit and JANA was used for magnetic structure refinement[31,32].

**Result and Discussion**

**A. Room temperature Structural properties**

Figure 2(a, b) shows the Rietveld refined XRD patterns of $GeCo_2O_4$ and GMCO recorded at room temperature. The targeted stoichiometric for Mn substituted compound was $Ge_{0.5}Mn_{0.5}Co_2O_4$. For both the compositions, all the observed peaks are very well fitted with a single cubic spinel structure with space group $Fd$-$3m$, consistent with the reported crystal structure of GCO and $Ge_{0.8}Mn_{0.2}Co_2O_4$[33]. In case of $Ge_{1-x}Mn_xCo_2O_4$, the refinement was performed assuming a random distribution of $Ge^{4+}$ and $Mn^{4+}$ in 50:50 at the tetrahedral site $8b$, $Co^{2+}$ at the octahedral site $16c$ and $O^{2-}$ at the site $32e$. The refined lattice parameters for GCO and GMCO are a = 8.3095(6) and 8.3085(6) Å respectively. The inset in each panel of figure 2 enlarges the (008) reflection exhibiting Cu- $K_{\alpha1}$ and $K_{\alpha2}$ splitting. No extra peak at the lower scattering angle was observed, which confirms that the Ge and Mn are randomly distributed at the 8b site. Also, no unaccounted peaks were observed in the data, suggesting the possibility of single-phase nature of both the samples. It is important to note that in a solid solution like $Ge_{0.5}Mn_{0.5}Co_2O_4$, accurately determining the Ge:Mn ratio using X-ray diffraction is challenging due to the relatively similar atomic scattering factors of Ge and Mn. While a difference of seven electrons in their atomic numbers provides some contrast, it is often insufficient in complex or disordered systems, where site mixing and partial occupancies can obscure compositional distinctions. Therefore, it remains challenging to conclusively identify phase inhomogeneity in the $Ge_{1-x}Mn_xCo_2O_4$ based solely on XRD data.

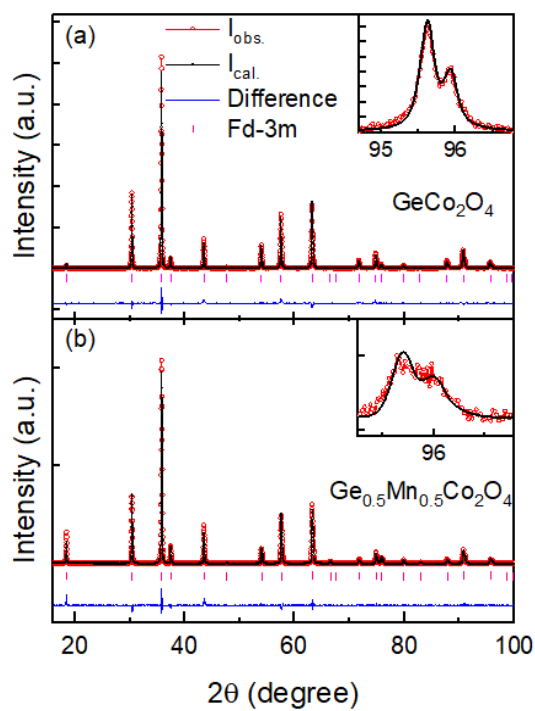

Figure 2: Rietveld refined room temperature XRD patterns of (a) GeCo$_2$O$_4$ and (b) Ge$_{1-x}$Mn$_x$Co$_2$O$_4$. These are refined using the *Fd*-3*m* spinel structure. The inset in each panel zooms in on the (008) reflection with Cu K$_{\alpha 1}$ and K$_{\alpha 2}$ splitting.

To better visualize the sample's compositional inhomogeneity, high-resolution imaging, Backscattered Electron X-ray (BEX), and Energy-Dispersive X-ray (EDX) analysis have been performed. The BEX layered and SE image given in figure 3(a,b) provide an overview of the sample, showing grain size distribution and compositional variations across different regions, with grain sizes generally below 5 µm and noticeable material clustering. The EDX spectra on powder sample over a broad area gives the stoichiometric ratio close to Ge$_{0.5}$Mn$_{0.5}$Co$_2$O$_4$ but the detailed EDX on epoxy mounted samples shows striking compositional inhomogeneities which are crucial to understand the magneto-structural properties of Mn substituted compound.

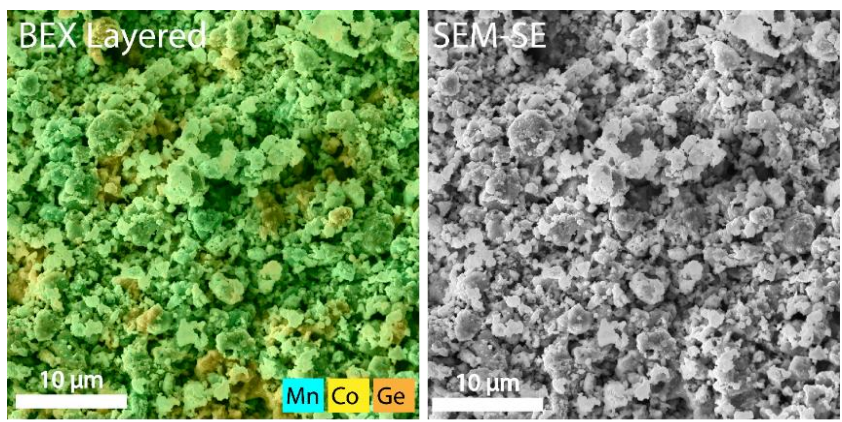

**Figure 3**: SEM-BEX Layered and SE image of the Mn-Ge-Co sample, illustrating overall microstructure, grain size, and chemical composition variations. with locations marked for Spectrum A and Spectrum B, where EDX spectra were collected. The spectra display variations in Ge/Mn ratios, corresponding to two distinguishable compositions. The measured values are given in the insert.

The SEM-EDX analysis of the epoxy-mounted sample shows two distinct compositional phases, marked as region A and B in figure 4. The BEX image shows a clear colour contrast between two phases: green (region A) and yellow (region B). Region A is Mn-rich, while Region B is Ge-rich, indicating the presence of two distinct compositional phases in the sample. Both regions exhibit compositions close to the AB$_2$O$_2$ stoichiometry, with Ge/Mn at A site and Co at B site; however, they differ in the A-site occupancy. The major phase (region A) exhibits a composition approximating Mn$_{0.74}$Ge$_{0.18}$Co$_2$O$_4$ (Figure 4, Spectrum A) whereas the Ge-rich secondary phase (region B) is closer to Ge$_{0.91}$Mn$_{0.18}$Co$_2$O$_4$ (Figure 4, Spectrum B). These results show that, despite XRD indicating a uniform stoichiometry, microscopic analysis reveals significant local compositional inhomogeneities. When phases have similar lattice parameters, standard laboratory XRD lacks the resolution to distinguish them, highlighting the need for high-resolution or complementary techniques.

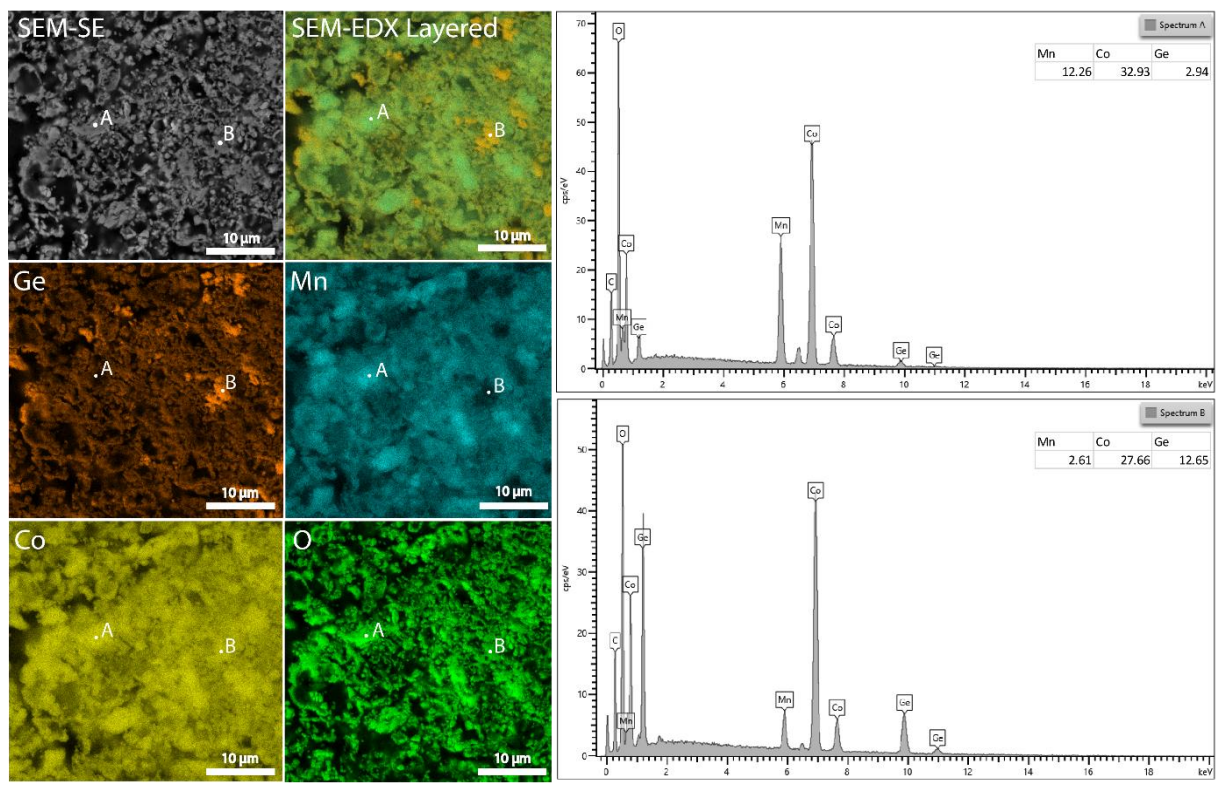

**Figure 4**: SEM-EDX analysis results on the epoxy mounted sample showing secondary electron (SE) imaging, EDX layered mapping, and elemental composition maps, with locations marked for Spectrum A and Spectrum B, where EDX spectra were collected. The spectra display variations in Ge/Mn ratios, corresponding to two distinguishable compositions. The measured values are summarized in the table.

However, due to the substantial difference in the neutron scattering lengths for Mn (-3.73 fm) and Ge (8.185 fm), neutron diffractions provide an advantage over XRD in $Ge_{1-x}Mn_xCo_2O_4$. The NPD pattern from the ECHIDNA beamline was analysed to check for the possible phase inhomogeneity in the as prepared samples. The initial model, which assumed both Ge(50%) and Mn(50%) at the A site, failed to capture the full intensity of several reflections. Even after several iterations and checking all the possibilities of cation distribution at the A and B sites, this model could not capture the full intensity of some of the reflections, as shown in figure 5a. To account for the full intensities of these reflections, an additional phase was added using the insight gained from SEM/EDX analysis. The corresponding refined pattern is presented in figure 5b. With the two-phase model, due to very similar cation distribution in these phases, the refinement was nontrivial. To ensure stability in the refinement, the chemical compositions of the two phases were fixed as $Mn_{0.74}Ge_{0.18}Co_2O_4$ and $Ge_{0.91}Mn_{0.19}Co_2O_4$, based on EDX data. The value of $\chi^2$ was reduced from 7.59 to 3.75, while the values of $R_{wp}$ reduced from 14.6 to 9.86 in compared to the fit given in figure 5a. The refined lattice parameters for the primary phase $Mn_{0.74}Ge_{0.18}Co_2O_4$ is $a$ = 8.29642(7) Å, while for the secondary phase $Ge_{0.91}Mn_{0.19}Co_2O_4$, the lattice parameter is slightly larger with $a$ = 8.30751(8) Å, by about 0.133%. The estimated phase fraction of the secondary Ge rich phase is 6.26(34) % and it is kept fixed for the refinement of the low-temperature NPD data for the subsequent sections. In the crystal structure of the main phase ($Mn_{0.74}Ge_{0.18}Co_2O_4$), the Mn/Ge atoms occupy the Wyckoff position 8$b$ at the coordinates (1/8, 1/8, 1/8), Co at 16$c$ (1/2, 1/2, 1/2) and the O atoms reside at 32$e$ site with fractional coordinates (z, z, z) where z = 0.2533(1). The isotropic mean square displacement parameter B for Ge/Mn, Co and O are 0.91(5), 0.65(5) and 1.10(1) Å$^2$. Due to the minute amount of the secondary phase, it was not possible to refine the displacement parameters for individual atoms in the secondary phase and the overall B factor was instead refined.

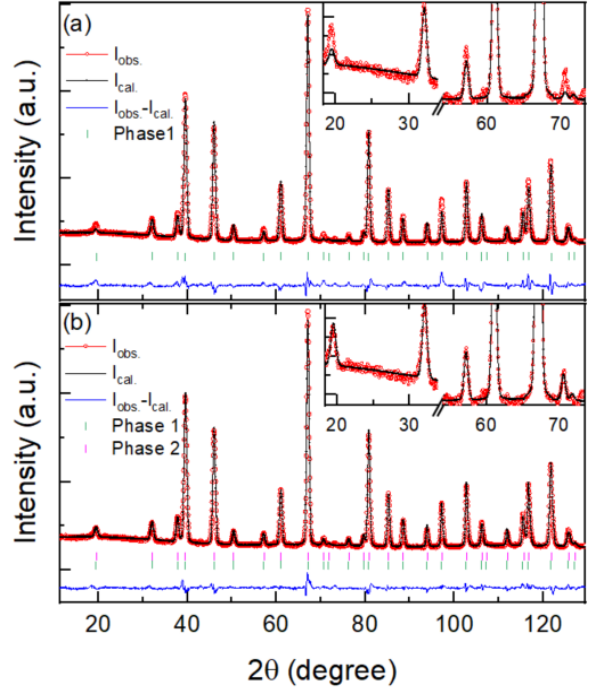

**Figure 5**: Rietveld refinement of room-temperature neutron powder diffraction (NPD) patterns for Ge$_{1-x}$Mn$_x$Co$_2$O$_4$ using (a) a single-phase model with x=0.5 and (b) a two-phase model with Mn$_{0.74}$Ge$_{0.18}$Co$_2$O$_4$ as the primary phase and Ge$_{0.91}$Mn$_{0.19}$Co$_2$O$_4$ as secondary phase. The insets highlight the 2θ range of 20°–70°, where noticeable discrepancies in peak intensities are observed for the single-phase model, while the two-phase refinement provides significantly improved agreement with the experimental data, indicating phase coexistence in the sample.

## B. Physical properties

To investigate the effect of Mn-substitution on the magnetic properties, magnetic susceptibility of Ge$_{1-x}$Mn$_x$Co$_2$O$_4$ has been measured and compared with GCO. Figure 6(a) shows the magnetic susceptibility χ(T) curves of both samples measured between 2 K and 300 K under zero-field-cooled (ZFC) and field-cooled (FC) conditions in an applied field of 500 Oe. The χ(T) of pristine GCO reveals a single AFM transition at $T_N$ = 22 K, consistent with the literature[3,18,34]. For GMCO, the ZFC χ(T) rises sharply below 120 K, showing a peak near 95 K with a clear bifurcation between ZFC and FC below it. The transition temperature, estimated from the first derivative, is $T_C$ = 108 K and is associated with the ferrimagnetic (FiM) ordering of A- and B-site spins in the Mn-rich main phase, as will be further confirmed by neutron diffraction data. At lower temperatures, a subtle kink appears around $T_N$ = 22 K, more prominently seen in the heat capacity data (Figure 6b). This second transition is attributed to the AFM ordering of Co spins in the Ge-rich phase. The isothermal magnetization curves are the typical curves expected from a ferrimagnetic lattice with large anisotropy below $T_C$. In MnCo$_2$O$_4$, the FiM order takes place below 184 K whereas the FiM transition temperature decreases to 77 K for Ge$_{0.8}$Mn$_{0.2}$Co$_2$O$_4$[24,33]. In the present case, with Mn$_{0.74}$Ge$_{0.18}$Co$_2$O$_4$ as the main phase, the transition temperature seems reasonable while comparing it with other members of this family. The inverse magnetic susceptibility in the inset of Figure 6(a) exhibits deviation from the Curie-Weiss (CW) behaviour below 200 K, which indicates that the magnetic interaction starts at a temperature far above T$_C$. Moreover, the value of Curie constant is 6.04 which corresponds to the effective paramagnetic moment value of 6.95($\pm$0.03) $\mu_B$/f.u. The value of Curie temperature ($\theta_C$) is negative and equals -107.4 K. Considering the cation distribution Mn$_{0.74}$Ge$_{0.18}$Co$_2$O$_4$, the spin only value of $\mu_{eff.}$ will be 6.40 for high spin and 4.12 $\mu_B$ for low spin states of Co$^{2+}$. In the absence of JT distortion (and associated structural transition), the observed value of $\mu_{eff.}$ suggests that the Co$^{2+}$ is in high spin state.

The heat capacity and magnetic entropy ($S_{mag.}$) of Ge$_{1-x}$Mn$_x$Co$_2$O$_4$ are presented in Figure 6(b). A broad transition with a peak centre at 108 K ($T_c$) is observed in the data, associated with the fiM order in the main Mn rich phase. An additional sharp peak at 22 K is also evident, linked with the AFM ordering in the secondary Ge rich phase. The broad nature of the high-temperature transition is attributed to the slightly varying Ge/Mn ratio (2-3%) across different regions of the sample, as confirmed by SEM/EDX. A combination of Debye and Einstein models is used to evaluate the lattice specific heat[35]. The estimated value of magnetic entropy changes equals 11.26 J/mole-K. Assuming Mn$^{4+}$ at A site and Co$^{2+}$ at B site, this value is nearly 56% of the expected value for Mn$_{0.74}$Ge$_{0.18}$Co$_2$O$_4$, consistent with the reported values for other members of this family. An alternative method based on harmonic lattice approximation was used by Lashley *et al.*[4] to estimate the magnetic entropy for GCO and GeNi$_2$O$_4$, which yield almost identical results. The reported values of magnetic entropy were only 58.3% and 56.5% in the case of GCO and GeNi$_2$O$_4$ and the missing entropy was suspected to be originating from substantial magnetic correlations well above T$_N$ [4].

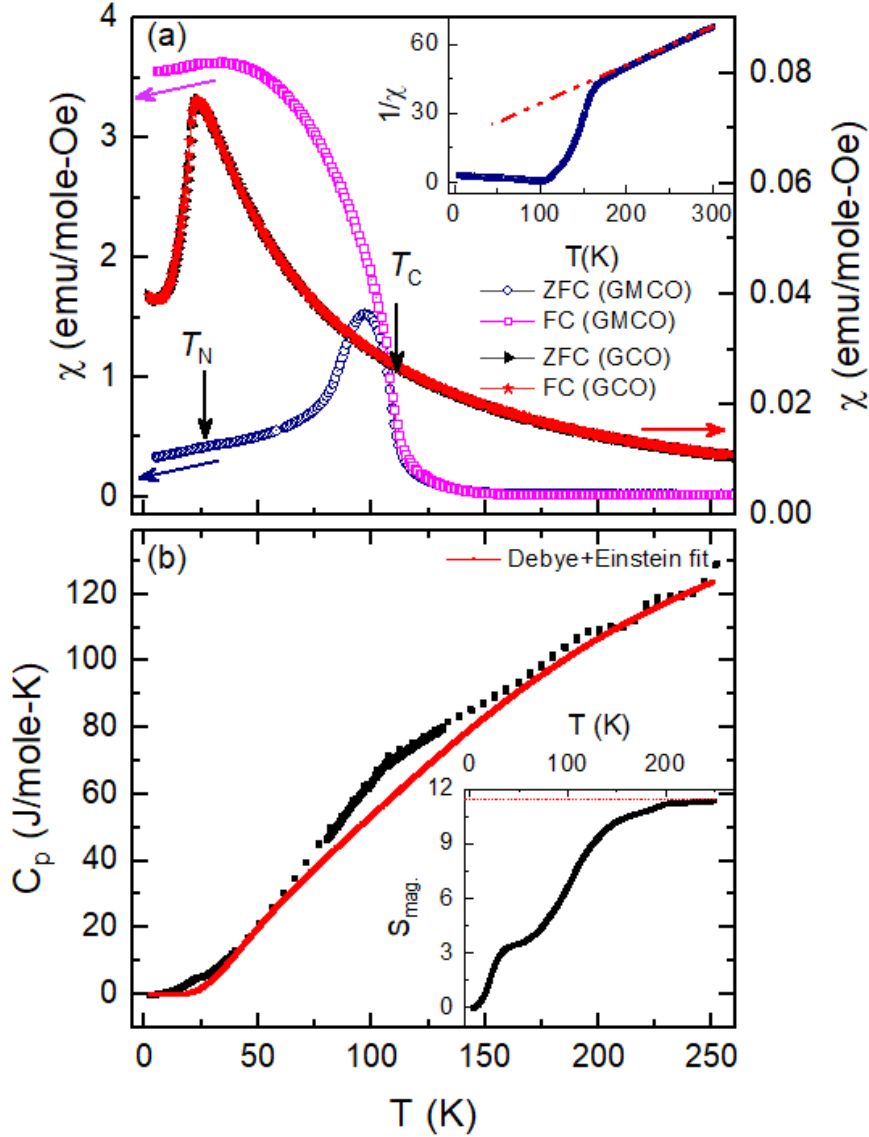

**Figure 6:** Temperature-dependent (a) DC magnetic susceptibility of GCO and GMCO and (b) heat capacity of GMCO. Inset in (a) shows the Curie–Weiss (CW) fit (red line) to the inverse susceptibility data for GMCO. Inset in (b) displays the magnetic entropy change associated with the two magnetic transitions. The red line in (b) represents the lattice contribution modelled using a combination of Debye and Einstein terms.

## C. Low temperature structural properties

To investigate the possible structural changes associated with AFM ordering, XRD patterns in the selected 2θ range were recorded for both GCO and $Ge_{1-x}Mn_xCo_2O_4$, while cooling the samples from 25 to 12 K. Figure 7(a, b) shows the temperature evolution of the (0 0 8) reflection. Clear splitting of the (0 0 8) reflection (both $K_{α1}$ and $K_{α2}$) at the cubic to tetragonal ($I4_1/amd$) transition in GCO, associated with the magnetic order is observed. Regarding the splitting in GCO, it can be seen starting from $T_N$ (= 21 K). Earlier, Barton et al.[19] argued that the structural transition in GCO is decoupled from the magnetic transition based on powder data. However, in the present case, it occurs at the same temperature, indicating a strong coupling between the structural and magnetic orders. Similar distortion at the Néel temperature in CoO has been under debate. While some reports indicate spin-orbit coupled magnetostriction[3,36] arise due to degenerate $t_{2g}$ states in octahedral $Co^{2+}$, others suggest JT ordering[37,38]. Interestingly, no such peak splitting or peak broadening effect was observed for $Ge_{1-x}Mn_xCo_2O_4$, indicating that it remains cubic down to 12 K. Furthermore, the full width at half maximum (FWHM) of (0 0 8) reflection remains almost invariant while cooling the sample. The estimated value of FWHM at 160 and 12 K are 0.322(1) and 0.319(9)°, respectively, confirming the absence of structural phase transition in $Ge_{1-x}Mn_xCo_2O_4$ within the resolution of the measurement. The origin of structural distortion in these systems with degenerate $t_{2g}$ states is usually the results of coupled effects of spin, orbital, and lattice degrees of freedom. Due to the complex cation distribution in the $Ge_{1-x}Mn_xCo_2O_4$ and several possible factors contributing to the distortions in these compounds, it is nontrivial to pinpoint a specific cause for the absence of a structural phase transition. Moreover, these distortions can be suppressed or decoupled from the magnetic ordering, resulting in the onset of AFM ordering without any accompanying structural transition [37]. Suppression of structural transition in the presence of minute Mn at the A site was also evident for the other members of this family[29].

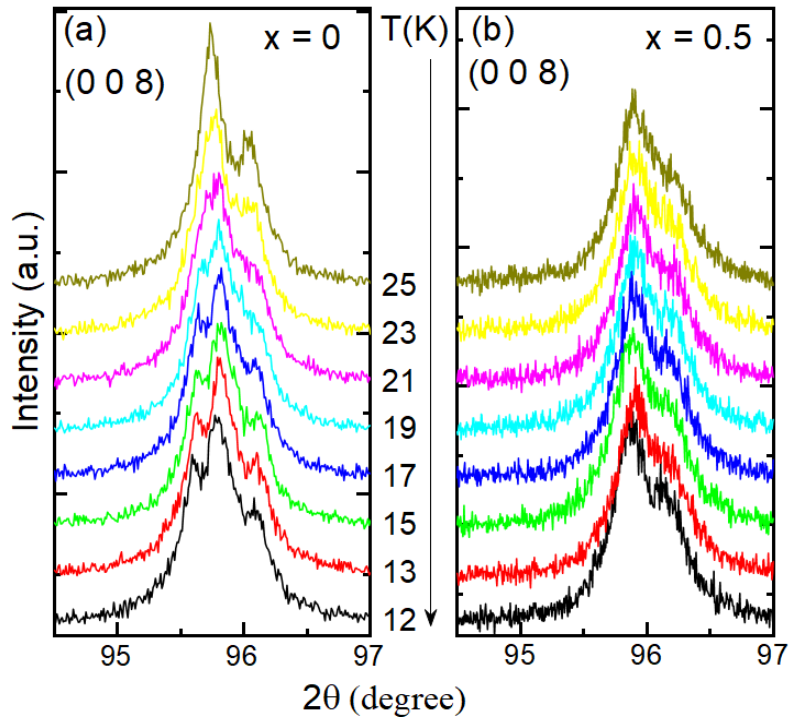

**Figure 7**: Temperature-dependent X-ray diffraction profiles of the (008) reflection for (a) GCO and (b) GMCO, illustrating the evolution of possible lattice distortion with decreasing temperature.

Figure 8(a-b) shows the Rietveld refined XRD patterns of GCO and $Ge_{1-x}Mn_xCo_2O_4$, collected at 12 K. The insets in each figure enlarged the view of (0 0 8) peak. The XRD pattern of GCO is refined using tetragonal symmetry $I4_1/amd$ whereas the pattern of $Ge_{1-x}Mn_xCo_2O_4$ has been refined using cubic symmetry, similar to the RT pattern. The refined lattice parameters at 12 K for GCO are 5.8777(1) and c = 8.3004(1) Å whereas for $Ge_{1-x}Mn_xCo_2O_4$, a = 8.29658(9) Å. The refined structural parameters for GCO at 12 K are given in table 2.

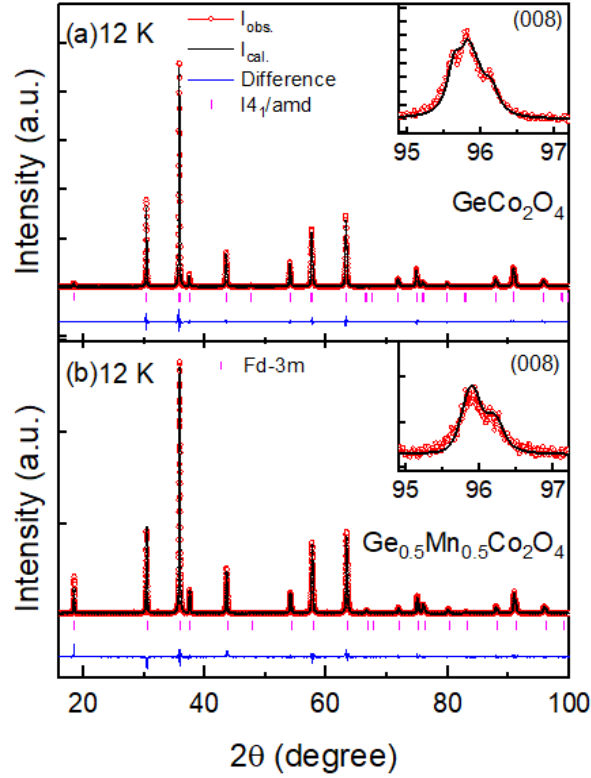

**Figure 8**: Rietveld refined XRD patterns of (a) GeCo$_2$O$_4$ and (b) Ge$_{0.5}$Mn$_{0.5}$Co$_2$O$_4$ measured at 12 K.

**Table 2**: Refined structural parameters of GCO at 12 K, obtained from Rietveld refinement using tetragonal symmetry ($I4_1/amd$). The refined lattice constants are $a$ = 5.8777(1) Å and $c$ = 8.3004(1) Å. Goodness-of-fit indicators: GoF = 1.61, $R$p = 7.13%.

| Atoms (site) | $x$ | $y$ | $z$ | B (Å$^2$) |
|---|---|---|---|---|
| Ge (8$b$) | 0 | 0.25 | 0.375 | 0.92(5) |
| Co (16$c$) | 0 | 0 | 0 | 0.65(5) |
| O (32$e$) | 0 | 0.5010 | 0.2519 | 1.10(1) |

**Neutron diffraction**

Figure 9(a) presents the NPD patterns at the selected temperature, measured with λ = 2.41 Å. The difference curves are also depicted in Figure 9(b) for selected temperatures. While cooling the sample below $T_C$, the intensity of some of the reflections at lower 2θ increases significantly, indicating a commensurate FiM order, associated with the propagation vector (0,0,0). The integrated intensity of the (1 1 1) reflection highlighted within the box region is plotted in the inset of Figure 9(b). As the temperature further decreases, several new peaks emerge adjacent to nuclear reflections below $T_N$, originating from the AFM ordering of Co spins in the secondary phase, Ge$_{0.91}$Mn$_{0.19}$Co$_2$O$_4$. These additional magnetic reflections, clearly visible in the difference data (30 – 5 K) shown in figure 9b, are associated with the propagation vector (1/2, 1/2, 1/2), analogous to the parent compound GCO[21,26,29].

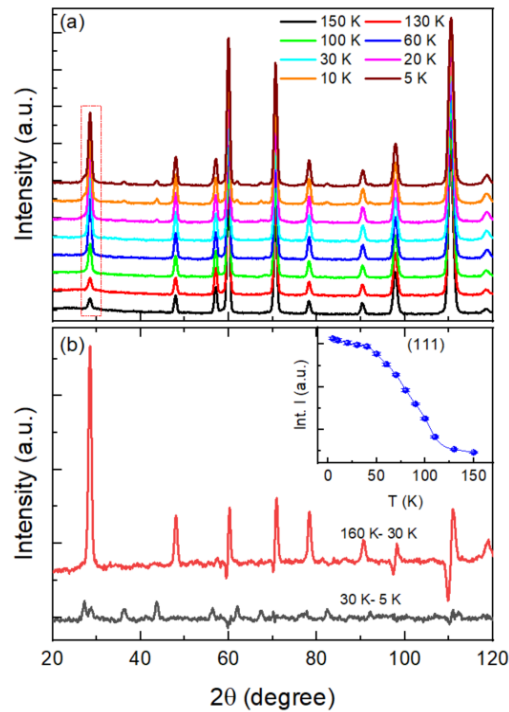

**Figure 9:** Neutron powder diffraction (NPD) patterns of $Ge_{1-x}Mn_xCo_2O_4$ recorded at selected temperatures. (a) Temperature-dependent diffraction patterns highlighting the emergence of magnetic Bragg peaks at low temperatures. (b) Difference patterns obtained by subtracting the high-temperature data, revealing two distinct sets of magnetic reflections associated with different magnetic sublattices or phases. The inset in (b) shows the temperature evolution of the (111) magnetic peak intensity.

To determine the magnetic structure of the sample above $T_C$ 160 K data was refined. For this purpose, two nuclear phases were modeled keeping their relative phase fraction fixed as obtained from the refinement of the RT Echidna data, see figure 2. Figure 10 shows the magnetic refinement of the data obtained below $T_N$, at 30 K and 5 K. Although the Ge rich phase is present in minute quantity, clear magnetic reflections can be observed adjacent to the nuclear reflections at 5 K. At 30 K, the strongest magnetic intensity is observed at the (111) reflection, indicating a ferrimagnetic order with propagation vector (0,0,0). We didn't observe any intensity at the location of (200) reflection, see the inset of figure 10b. This suggests complete absence of any additional AFM component, unlike $Ti_{1-x}Mn_xCo_2O_4$, where a weak intensity of (200) reflection is evident [30]. To solve the magnetic structure of the Mn-rich main phase, irreducible representations (IRREP) were calculated using the BASIREPS program with Mn at (1/8,1/8,1/8) and Co at (1/2,1/2,1/2) site. Out of the 10 possible IRREPs, the data could be fitted well only with Γ8 of dimension 3 contained only 1 time in Γ8. During the refinement, the occupancies and scale factor were kept fixed identical to that of 160 K refinement. The AFM coupled moments at A and B site exhibits collinear ferrimagnetic ordering, where the moments are aligned along the [001] and [00-1] directions. Furthermore, the absence of the magnetic (2 0 0) cubic peak excludes the Yafet-Kittel (Y-K) type spin-canted structure and supports the collinear magnetic structures. The $R_{wp}$ and $\chi^2$ values at 30 K are 7.48 and 6.15, respectively. The resultant magnetic structure of $Mn_{0.74}Ge_{0.18}Co_2O_4$ is shown in Figure 11(a), whereas the temperature variation of the A-site ($Mn_A$) and B-site ($Co_B$) moments are shown in in Figure 11(b). As the sample is cooled, both the $Mn_A$ and $Co_B$ moment size increases and attain the maximum value of around 20 K, which remains almost constant invariant while further cooling down to 5 K. At 30 K, the Mn moment is 2.30(4) $\mu_B$ and the Co moment is 1.79(3) $\mu_B$, which is consistent with the values of A and B site moment in similar systems[29,30].

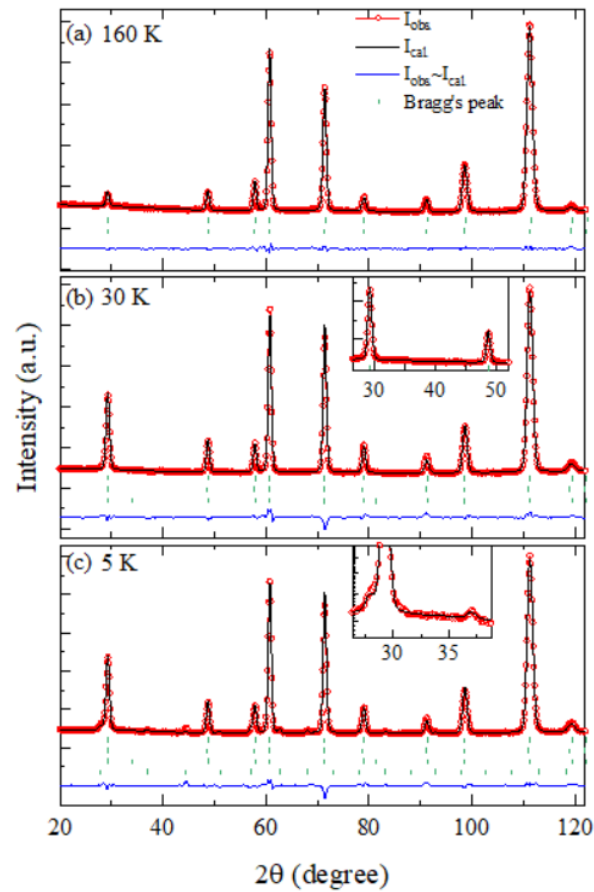

**Figure 10**: Neutron powder diffraction patterns of $Ge_{1-x}Mn_xCo_2O_4$ measured at (a) 160 K (above the magnetic ordering), (b) 30 K and (c) 5 K. In (a) two sets of nuclear Bragg peaks from the primary and secondary crystalline phases are visible. Panels (b) and (c) each show nuclear (upper) and magnetic (lower) reflections. Inset in (b) (around $2\theta = 34°$) confirms the absence of the (200) peak, while inset in (c) highlights the low-angle magnetic reflections associated with Co ordering in the secondary phase.

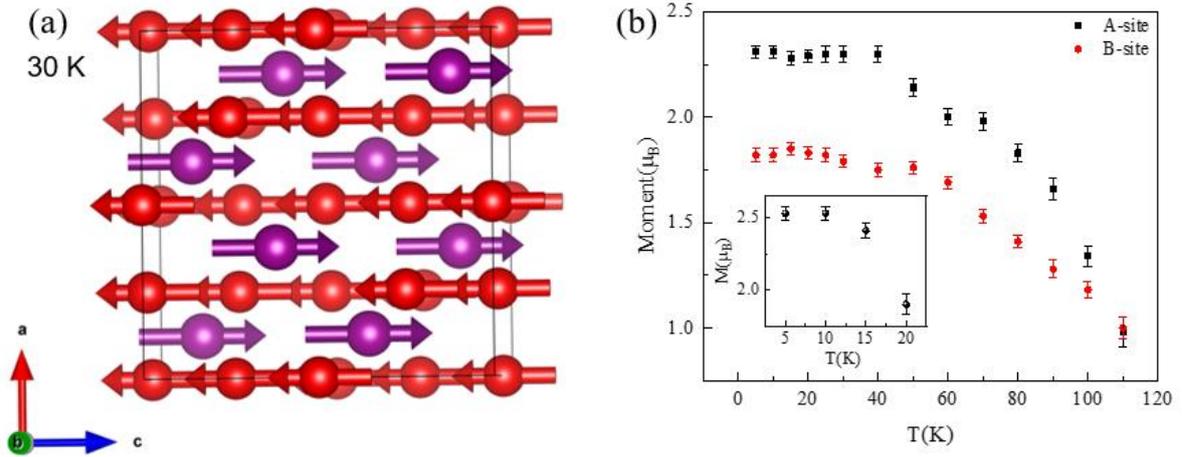

**Figure 11**: (a) Refined magnetic structure at 30 K, with A-site and B-site magnetic ions shown as red and violet spheres and arrows, respectively. (b) Temperature dependence of the average magnetic moments on the A-site (red) and B-site (violet) sublattices. Inset in (b) shows the temperature evolution of the Co moment in the secondary phase.

For the magnetic structure refinement below 25 K associated with AFM ordering of Co spins in Ge rich minor phase, an additional magnetic phase with propagation vector (½, ½, ½) was added in the input file (.pcr) to account for the extra magnetic peaks appearing below 22 K. The resultant magnetic structure in this minor phase resembles the parent compound GCO[12,18]. Also, for the refinement of the secondary phase, the scale factor is kept the same as that of the second nuclear phase. The average B site moment at 5 K in this secondary phase is 2.53(5) $\mu_B$ which is consistent with the parent compound[30]. The inset in Figure 11b exhibits the temperature dependence of refined Co moment in the minor Ge rich phase. The Mn atoms in this Ge-rich phase does not participate in the magnetic ordering below 22 K.

**Conclusion**

A comprehensive study of the crystal and magnetic structure of $Ge_{1-x}Mn_xCo_2O_4$ has been conducted, revealing the effects of Mn substitution on magneto-structural properties. The findings were compared with those of $GeCo_2O_4$ to understand the influence of added disorder at A site. SEM-EDX analyses in Mn substituted sample reveal nanoscale compositional inhomogeneities, providing crucial insights into phase separation and local stoichiometry that are not captured by bulk techniques. The GMCO sample predominantly consists of two distinct spinel phases, differentiated by A-site occupancy: a Mn-rich primary phase $Mn_{0.74}Ge_{0.18}Co_2O_4$ and a minor Ge-rich secondary phase $Ge_{0.91}Mn_{0.19}Co_2O_4$. Two-phase Rietveld refinement of room-temperature neutron powder diffraction data confirms the presence of minor secondary phase $Ge_{0.91}Mn_{0.19}Co_2O_4$ with a phase fraction of approximately 6.26(34)%. Both phases adopt a cubic spinel structure, exhibiting only subtle differences in lattice parameters, with values of a = 8.29642(7) Å for the Mn-rich phase and a = 8.30751(8) Å for the Ge-rich phase. Comparison of low temperature structural properties of GCO and GMCO exhibits striking differences, despite having nearly identical structure at room temperature. The low-temperature XRD results confirm a cubic-to-tetragonal transition in GCO associated with magnetic order, which remains absent in GMCO. This could be attributed to the added disorder at the A site in GMCO which is possibly suppressing the structural distortions linked with magnetic transitions. Similar effects have been earlier observed for other Mn substituted compounds[24,30]. Ferrimagnetic ordering of Mn and Co spins is confirmed in the main Mn rich phase below 108 K, while the Ge-rich phase exhibits AFM ordering of Co spins below 22 K. The refined magnetic moment values in both phases of GMCO are consistent with reported literature values. Our extensive analyses using SEM-BEX, EDX, X-ray, and neutron diffraction reveal nanoscale compositional inhomogeneities, providing crucial insights into phase separation and local stoichiometry, which are essential for understanding the intrinsic magnetic response of these compounds.

**Acknowledgements**

A portion of this work was performed at the National High Magnetic Field Laboratory, which is supported by National Science Foundation Cooperative Agreement No. DMR-2128556 and the State of Florida. B.S. and K.W. acknowledge the support of the NHMFL User Collaboration Grant Program. SEM-BEX and EDX work was performed at the Center for Rare Earths, Critical Minerals, and Industrial Byproducts at the National High Magnetic Field Laboratory, Florida State University, supported by the State of Florida through Contract # 0000071627. T.S. acknowledges funding from the National Science Foundation under Grant No. DMR-2219906